\documentclass[twocolumn]{webofc}

\usepackage[varg]{txfonts}   
\begin{document}
\title{A Scintillator and Radio Enhancement of the IceCube Surface Detector Array}
%
%

\author{\firstname{Andreas} \lastname{Haungs}\inst{1}\fnsep\thanks{\email{andreas.haungs@kit.edu}} 
        \lastname{for the IceCube Collaboration}
}

\institute{Karlsruhe Institute of Technology - KIT, Institut f\"ur Kernphysik, 76021 Karlsruhe, Germany}

\abstract{%
  An upgrade of the present IceCube surface array (IceTop) with scintillation detectors and possibly radio antennas is foreseen. The enhanced array will calibrate the impact of snow accumulation on the reconstruction of cosmic-ray showers detected by IceTop as well as improve the veto capabilities of the surface array.  In addition, such a hybrid surface array of radio antennas, scintillators and Cherenkov tanks will enable a number of complementary science targets for IceCube such as enhanced accuracy to mass composition of cosmic rays, search for PeV photons from the Galactic Center, or more thorough tests of the hadronic interaction models. Two prototype stations with 7 scintillation detectors each have been already deployed at the South Pole in January 2018.  These R\&D studies provide a window of opportunity to integrate radio antennas with minimal effort.
}
\maketitle
\section{Introduction}
\label{intro}

The IceTop surface array of the IceCube Neutrino Observatory~\cite{detector} measures cosmic 
rays in the transition region from galactic to extra-galactic sources. 
However, the non-uniform snow accumulation on the 
installed ice-Cherenkov tanks leads to a non-uniform signal 
attenuation~\cite{icetop1, icetop2} 
which results in an increased uncertainty
on the reconstruction of the air-shower parameters.

An upgrade of IceTop with an array of scintillator panels is under consideration~\cite{scintseckel, scinthuber}.
The proposed enhancement foresees the deployment of 37 stations of 7 detectors each within the 
present IceTop area. Taking advantage of the infrastructure that the scintillator array 
will provide, installation of radio antennas is also being considered.
This requires only moderate additional effort and will transform the surface array into a 
multi-component detector that would also serve as platform for technology tests and for the design of a 
possible future multi-messenger large-scale observatory at the South Pole. 
In addition, the collaboration has examined the possibility to add Cherenkov telescopes (IceAct) 
to the surface instrumentation~\cite{iceact}.
In this paper we discuss the possible science benefits for a hybrid scintillator-radio instrumentation 
and show the status and efforts in prototyping the proposed IceTop enhancement. 

\section{Motivation and Science Case}

The proposed detector types are designed to serve the following general goals:
\begin{itemize}
\item Cross-calibration: The coincident detection of cosmic-ray induced air showers and muons deep in the ice will allow for an improved calibration of the in-ice detector and IceTop.
\item Improved capabilities for studying cosmic rays: The detection of comic rays through several independent detection channels will enhance the capabilities of IceCube and IceTop to measure the mass composition of cosmic rays. 
\item Lowering the threshold for air-shower observations: The higher density of detectors will allow accurate reconstruction of air showers for energies below 1 PeV, i.e. the full energy range of the knee will be covered.
\item Extending the sensitive area: Although not occupying a larger area, the new detectors will allow for analyzing cosmic ray data where the core lands outside the IceCube footprint.  
\item Better understanding of hadronic interaction models: The measurements of air showers through several detection channels will improve the understanding of hadronic interactions and thus the muon background for the neutrino measurements.
\item Improved sensitivity to primary gamma rays: For the gamma-ray detection of sources a larger energy range and a larger phase space will be available.
\item Improvement of surface veto capabilities for neutrinos: The energy threshold for vetoing the muon background to astrophysical neutrinos in IceCube will be reduced. 
\item R\&D effort towards a large surface extension of IceCube-Gen2: The scope of such an upgrade includes  
developmental steps towards a Gen2-type larger surface detector. 
\end{itemize}
\begin{figure}[t]
\centering
\vspace*{0.3cm}
\includegraphics[width=0.95\columnwidth,clip]{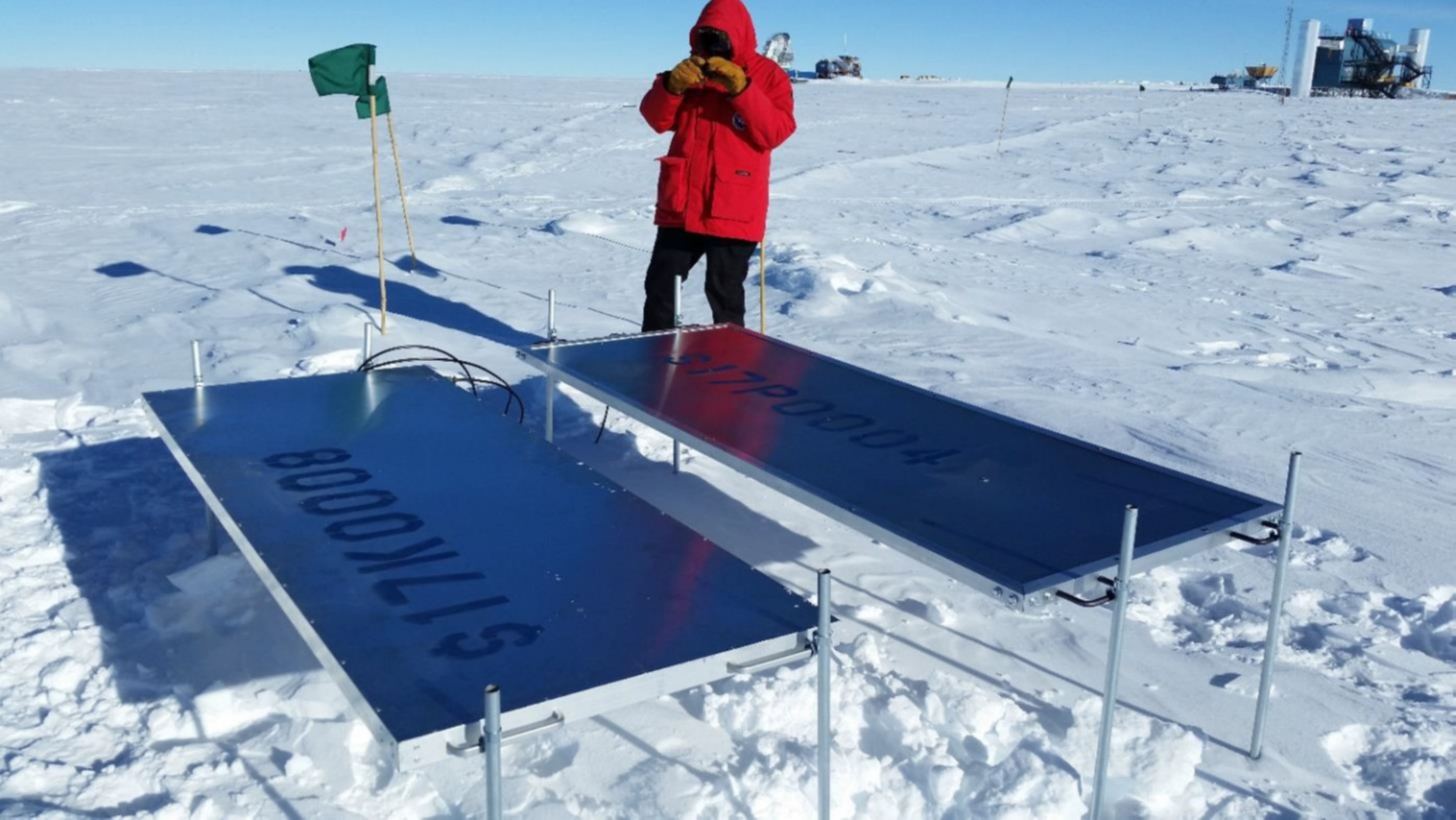}
\caption{Installation of the first prototype scintillation detectors at the South Pole in January 2018.}
\label{fig-1}       
\end{figure}

In particular, the {\bf scintillator} upgrade of IceTop will allow us to: \\
- Measure the effect of snow on the IceTop tank sensitivity, binned by energy, zenith, and radial distance from the shower core. This will calibrate performance of IceTop and reduce the current systematic uncertainty in the cosmic ray mass composition measurement. \\
- By doubling the active sensor area, scintillators significantly lower the detection threshold for cosmic rays permitting studies across the knee region.
\\
- Increase the muon veto capabilities for IceCube.  By adding scintillators with a similar coverage as IceTop, the energy threshold at which the veto becomes efficient at a 10$^{4}$ to 10$^{5}$ rejection factor is estimated to be lower by a factor of two. \\
- Combining scintillator, IceTop, and in-ice measurements will increase the accuracy and the energy range for cosmic-ray energy and mass composition measurements. Scintillators and IceTop tanks have different responses to the electromagnetic and muonic shower components, so the combination can disentangle their relative contribution in individual air showers~\cite{auger}. The electron-muon-ratio provides a strong parameter for the determination of the mass composition of cosmic rays. Thus, the scintillator array will improve the cosmic-ray and gamma-ray search performance of IceCube.  \\
An even higher quality of the multi-messenger capabilities will be reached by adding another independent shower detection technique, e.g. radio for the high-energy range or air-Cherenkov for the low-energy region. \\

{\bf Radio antennas} will be used for high-accuracy measurements of the size and depth 
($X_\mathrm{max}$) of the electromagnetic shower component with a 24/7 duty cycle. 
This provides a complementary energy and composition measurement with low systematic uncertainties 
boosting the cosmic-ray reconstruction accuracy of IceCube~\cite{ARENA2018}: \\
- The flux and mass composition of the primary cosmic rays as a function of the energy are classical
measures in cosmic-ray science. Features in these observables carry information on the propagation
and the origin of the cosmic rays. Due to the lower systematic uncertainties of the radio technique compared to surface particle detection, the surface antenna array will increase the per-event precision and the absolute accuracy in the reconstruction. Hence, by such a radio antenna array the
energy scale of IceTop can be compared to other cosmic-ray experiments with a radio extension~\cite{kastun}. \\
- Weak anisotropies in the all-particle flux of primary cosmic rays have been discovered below
2 PeV and above 8 EeV~\cite{anisoKG} . The maximum energy of cosmic rays produced in the Milky
Way is presumed to be in this intermediate energy range. However, the arrival directions in this energy
range are remarkably isotropic, possibly because the different anisotropies of the galactic and
extragalactic cosmic rays overlap. Since the extragalactic cosmic rays below $10^{19}\,$eV seem to
be mostly protons or alpha particles, but the galactic cosmic rays in this energy range
feature a significantly heavier composition, this puzzle can be solved by an 
event-to-event separation of light and heavy cosmic rays. While the scintillator 
extension will provide
this separation for nearly vertical showers, the radio extension will provide it for more inclined events. \\
- The separate measurement of the electromagnetic component by the radio array, of the electrons
and low-energy muons by the particle detectors at the surface, and of the high-energy muons in the
ice will enable more thorough tests of hadronic interaction models. Improving the knowledge of
air-shower physics at multi PeV energy, and in particular understanding the muon deficits in the
models, is important to better quantify the muon background in neutrino measurements by IceCube. \\
- Gamma-ray induced showers emit a slightly stronger radio signal than showers initiated by cosmic-ray
nuclei and contain an order of magnitude fewer muons. Thus, the combination of radio and muon
detectors is ideal to search for photons. 
The Galactic Center - a promising candidate source of PeV photons - is continuously visible at 61 degrees zenith angle from the South Pole. \\
- In addition, a radio surface array will serve as an R\&D area for next generation
projects, like IceCube-Gen2. 
In particular, surface antennas could be used to eliminate cosmic ray transients and 
anthropogenic backgrounds to a future in-ice radio detector of UHE astrophysical and/or 
cosmogenic neutrinos. \\
- Due to the large radio footprint of inclined showers, a radio surface array could
veto very inclined events, even if the shower core is far outside of the array. 
\begin{figure*}[t!]
\centering
\includegraphics[width=0.7\textwidth,clip]{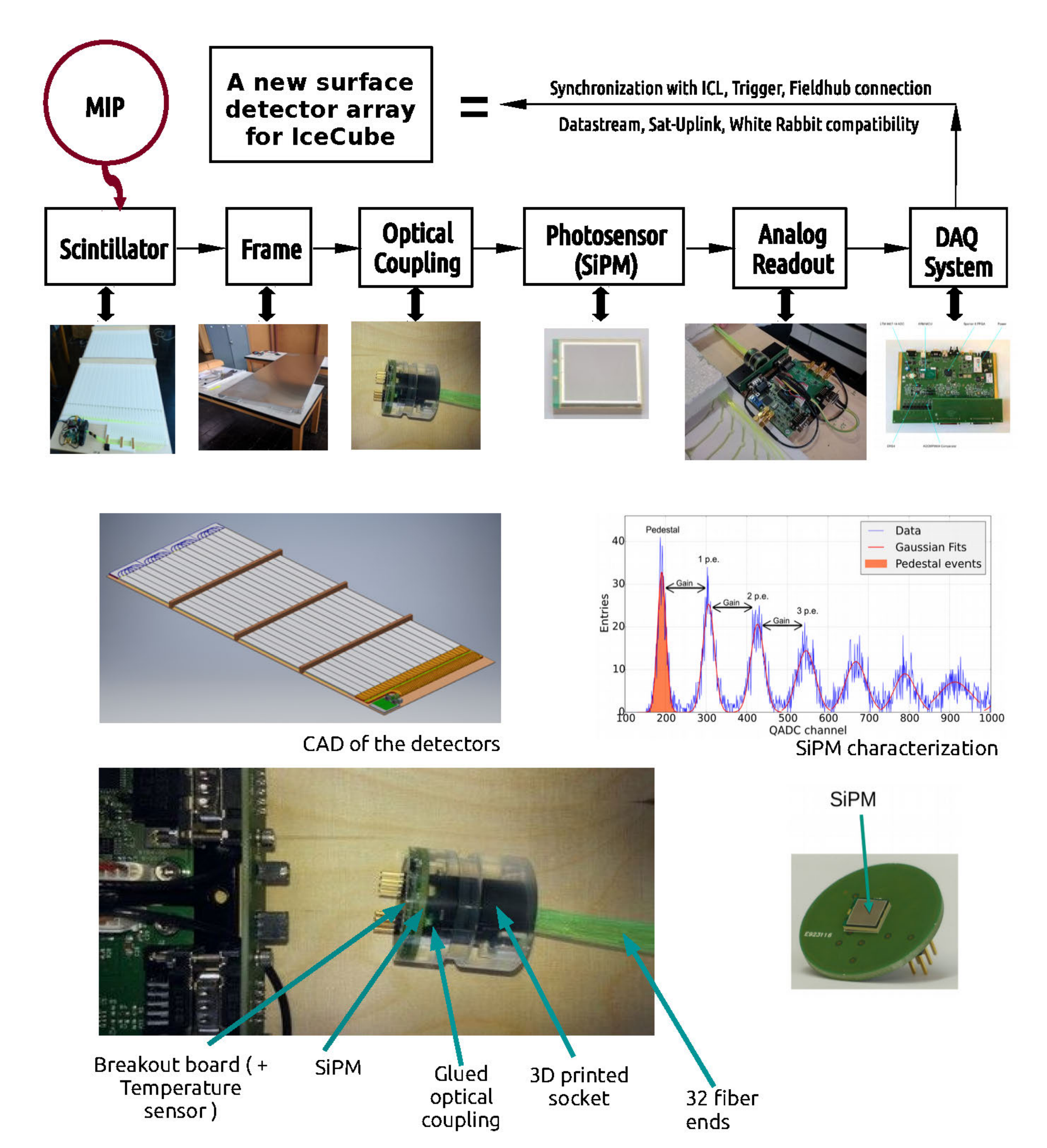}
\caption{Prototype scintillator detector design, here of the IceTAXI system. Taken from~\cite{huberTevpa}.}
\label{fig-2}       
\end{figure*}

\section{Prototyping}

\begin{figure}[t]
\centering
\includegraphics[width=0.9\columnwidth,clip]{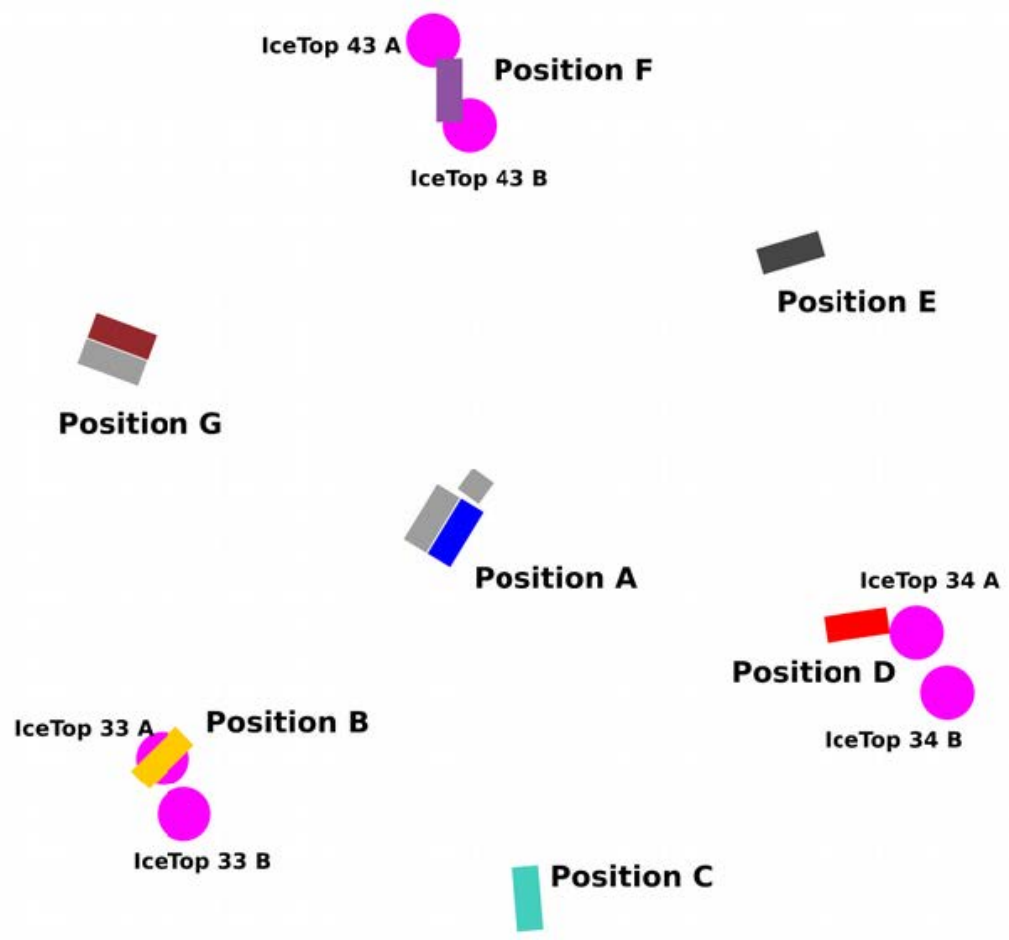}
\caption{Layout of the prototype scintillator stations at the South Pole around the ice-Cherenkov stations 33, 34, and 43. Partly, the scintillators are installed one above the other, partly side by side. The schematic is not-to-scale, but the distance between the scintillator positions is roughly $60\,$m.}
\label{fig-3}       
\end{figure}
\begin{figure}[h]
\centering
\includegraphics[width=0.9\columnwidth,clip]{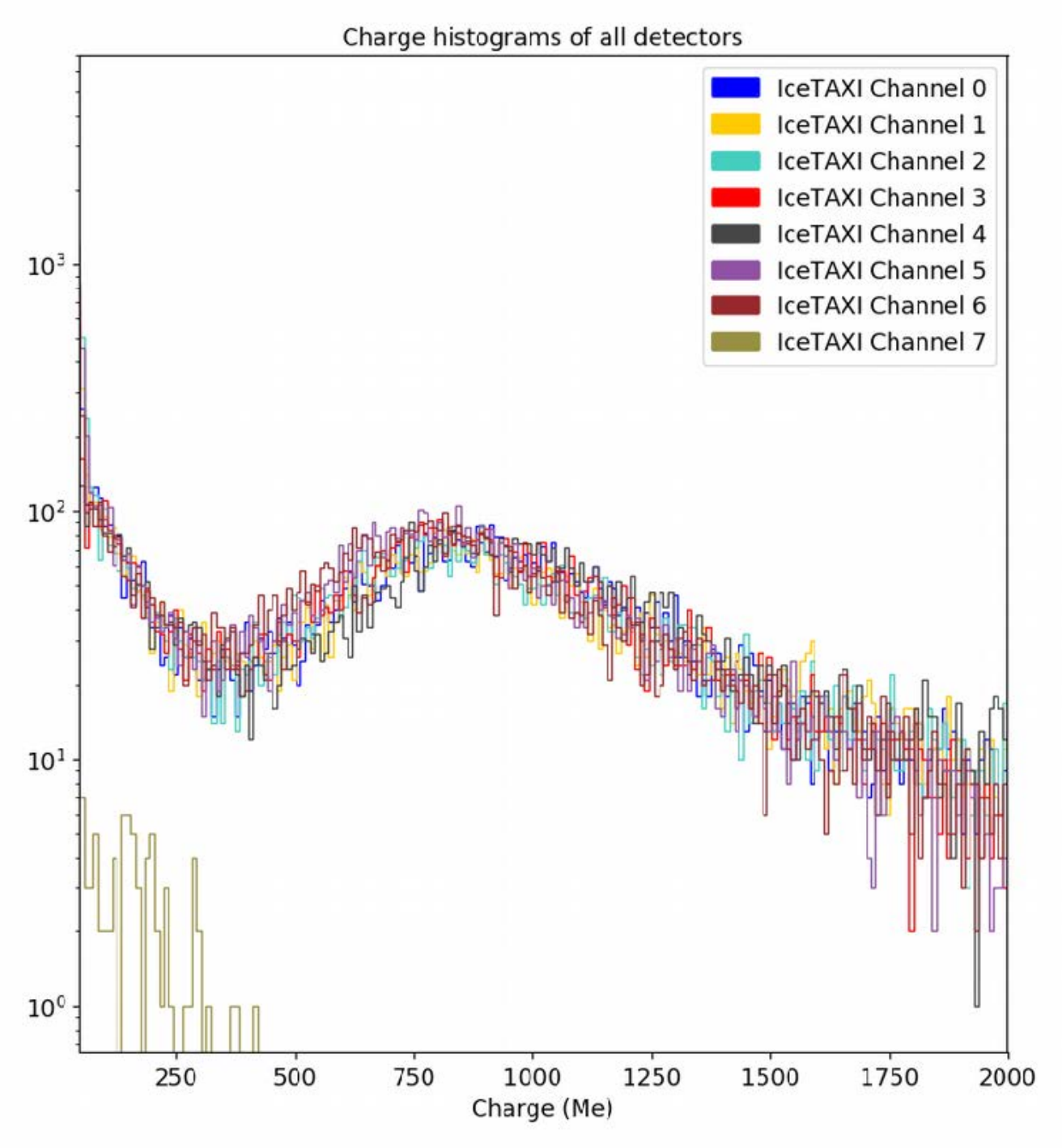}
\caption{Distribution of MIP particles of seven detectors at the South Pole. Channel 7 of the IceTAXI readout system is not equipped with a detector, i.e. intrinisc noise of the TAXI board is recorded, only.}
\label{fig-4}       
\end{figure}
In order to show the performance of scintillators within the IceTop area two prototype stations have been 
developed and installed at the South Pole in early 2018~\cite{phd-th} (Figure~\ref{fig-1}). 
A station comprises 7 scintillation detectors 
of 1.5$\,$m$^2$ sensitive area connected together and with the IceCube main building via a field hub box.
The scintillation panels are arranged in a hexagon with a distance of 60 m to each other.
The scintillators are readout by $6 \times 6\,$mm$^2$ Hamamatsu SiPMs and constructed similarly for both stations.   
The two stations are based on different DAQ systems with the main difference that in one case the 
digitization of the signals is done within the scintillator panel ($\mu$daq), and in the other case centrally with IceTAXI at the field hub. 

To enable direct comparisons of the two systems, the two stations are placed at the same location. 
Figure~\ref{fig-2} sketches the design and readout concept of the scintillator panels.
Figure~\ref{fig-3} give the layout of the two prototype stations at the South Pole. 
First measurements have shown very similar and very promising perfomances of both systems. 
Figure~\ref{fig-4} gives the MIP distributions of all panels of the IceTAXI station for a run of few minutes.    \\ 

\begin{figure}[t]
\centering
\includegraphics[width=0.8\columnwidth,clip]{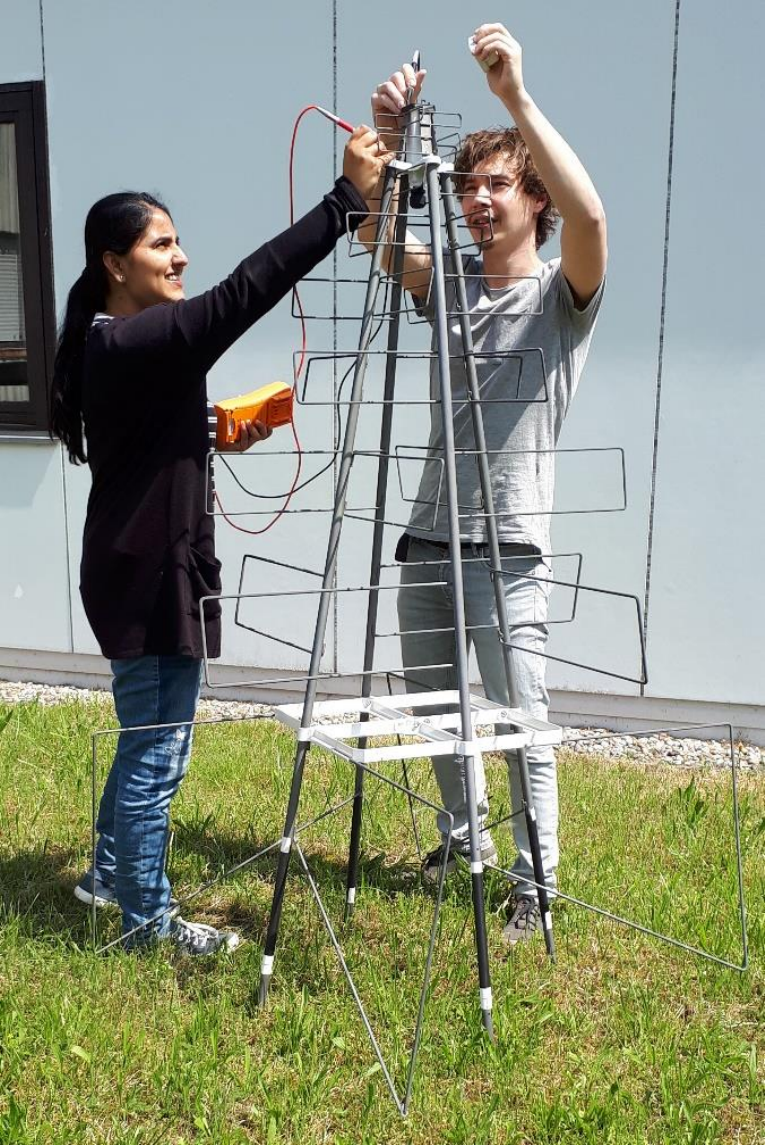}
\caption{Photo of a SKALA antenna installed at KIT for testing purposes.}
\label{fig-5}       
\end{figure}
Whereas for the scintillator part the system is now being optimized and prepared for mass production~\cite{phd-mo}, the 
integration of radio antennas is being further developed and first prototype antennas will be installed at the 
South Pole in early 2019.  
The radio array can be realized in a cost-effective way by docking and integrating antennas to the planned scintillator array. The idea is to share the infrastructure (cabling) and the data-acquisition~\cite{phd-mr}. 
By connecting the antennas to the same electronics, their readout
can easily be triggered by the scintillator detectors.  
One of the two DAQ options is the IceTAXI
system using three DRS4 chips for digitization of any detector signals~\cite{taxi}. Since only one DRS4
chip is needed for each station of seven scintillators, the remaining two chips can be used to connect
the antennas. 
The IceTAXI board houses also a FPGA, which enables to build a local trigger from the scintillators, also in case of the $\mu$DAQ readout system. 
Thus two or three antennas are planned per station, located close to the field hub. 
A good candidate for the antenna is the SKALA antenna~\cite{skala} developed for the
low-frequency array of the SKA telescope (Figure~\ref{fig-5}). 
SKALA features a low system noise and a smooth gain
pattern with a high sensitivity for the complete zenith-angle range of interest. First tests at
KIT show that the integrated low-noise amplifier operates well at the low temperatures expected at
the South Pole. Thus, we have chosen to install two SKALA antennas at the South Pole to
perform first prototype measurements triggered by the already operating scintillator prototypes.
The hybrid particle and radio DAQ is presently under development.

\section{Simulations of the scintillator array}

\begin{figure}[t]
\centering
\includegraphics[width=0.9\columnwidth,clip]{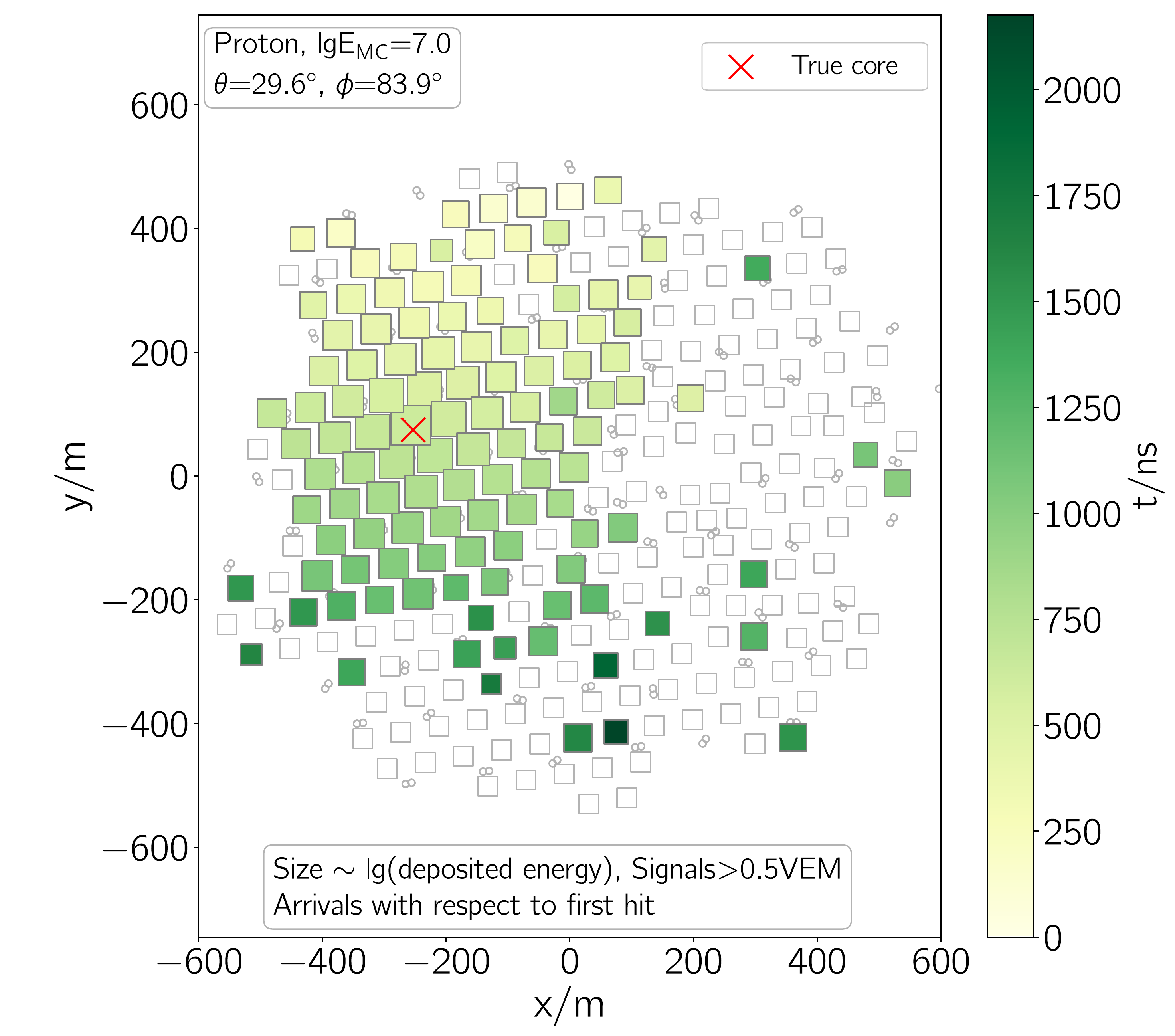}
\caption{Layout of the scintillator array used in the simulations. The layout is overlayed with the detector response to an individual shower, where the size of the squares denote the amplitude and the color the relative detection time.}
\label{fig-6}       
\end{figure}

In parallel to the hardware developments a detailed study of possible performances are studied via 
Monte-Carlo simulations~\cite{phd-al}. For the scintillator array we simulate the proposed enhancement of 37
stations within the IceTop area, where one station comprises 7 detectors (Figure~\ref{fig-6}). 
The chain of these simulations~\cite{agnieszkaecrs} combines the air-shower cascades and the 
interactions within the individual detectors. In a first step the scintillator array has been 
considered as an individual stand-alone detector. 
In a second step the information will be combined with the other types of detectors within the 
IceCube observatory.  

Extensive air showers are simulated with CORSIKA v7.5600 using Sybill 2.3 as high-energy
hadronic interaction model and FLUKA for the low energy interactions and an appropriate 
atmospheric model for the South Pole. 
The particles are read out at the observation level of $2838\,$m and further propagated to the 
scintillation array. 
The simulations were performed for proton and iron primaries of energies from 100 TeV to 100
PeV with incoming zenith angles from 0 to 40 degree and random azimuth, and several times 
resampled within the area of the array.
The implemented geometry of the detectors treats the scintillator bars as the
sensitive boxes coated with reflection layer and placed in an aluminum volume with styrofoam and
plywood. The simulations does not include the fibers and the fiber holes. The output of the
detector simulation is the energy deposited within the sensitive material. 
The simulated signal is defined as a sum of photons generated by all the particles crossing the detector. 
This unit is normalized to the number of photons generated by a vertically passing muon of 3 GeV energy 
(VEM = vertical equivalent muon). 

\begin{figure}[b!]
\centering
\includegraphics[width=0.9\columnwidth,clip]{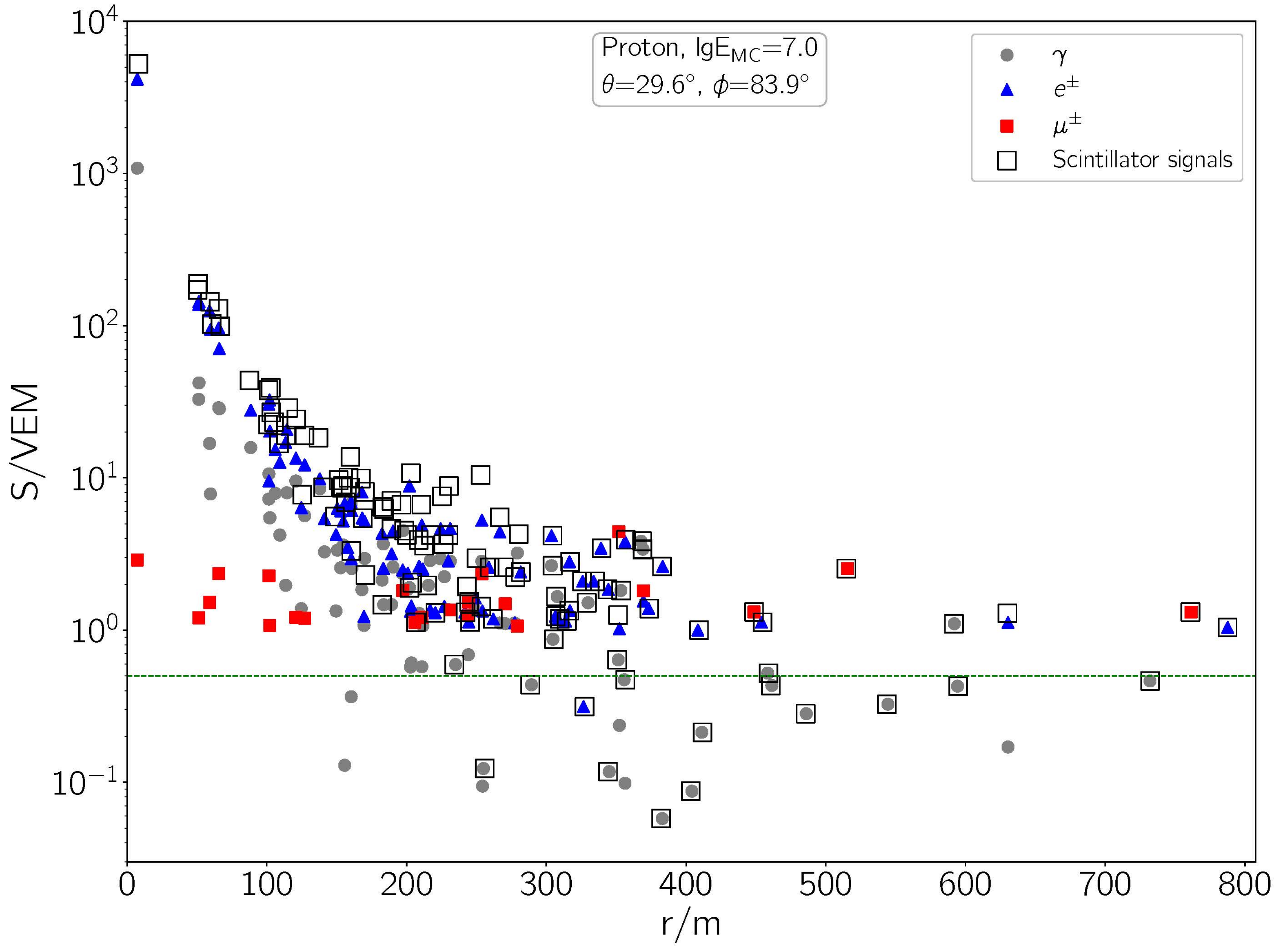}
\caption{Lateral distribution of air-shower
signals. The contributions from different
particle species are shown in addition.}
\label{fig-7}       
\end{figure}

Figure~\ref{fig-6} shows an example of the simulated response of the scintillation array to an air-shower
induced by a high-energy proton used for the reconstruction of the events.
Figure~\ref{fig-7} displays the lateral density distribution of a single event, where the
contributions from different particle species indicate the origin of the signals at different distances 
to the shower axis~\cite{agnieszkaecrs}. 

\begin{figure}[h]
\centering
\includegraphics[width=0.9\columnwidth,clip]{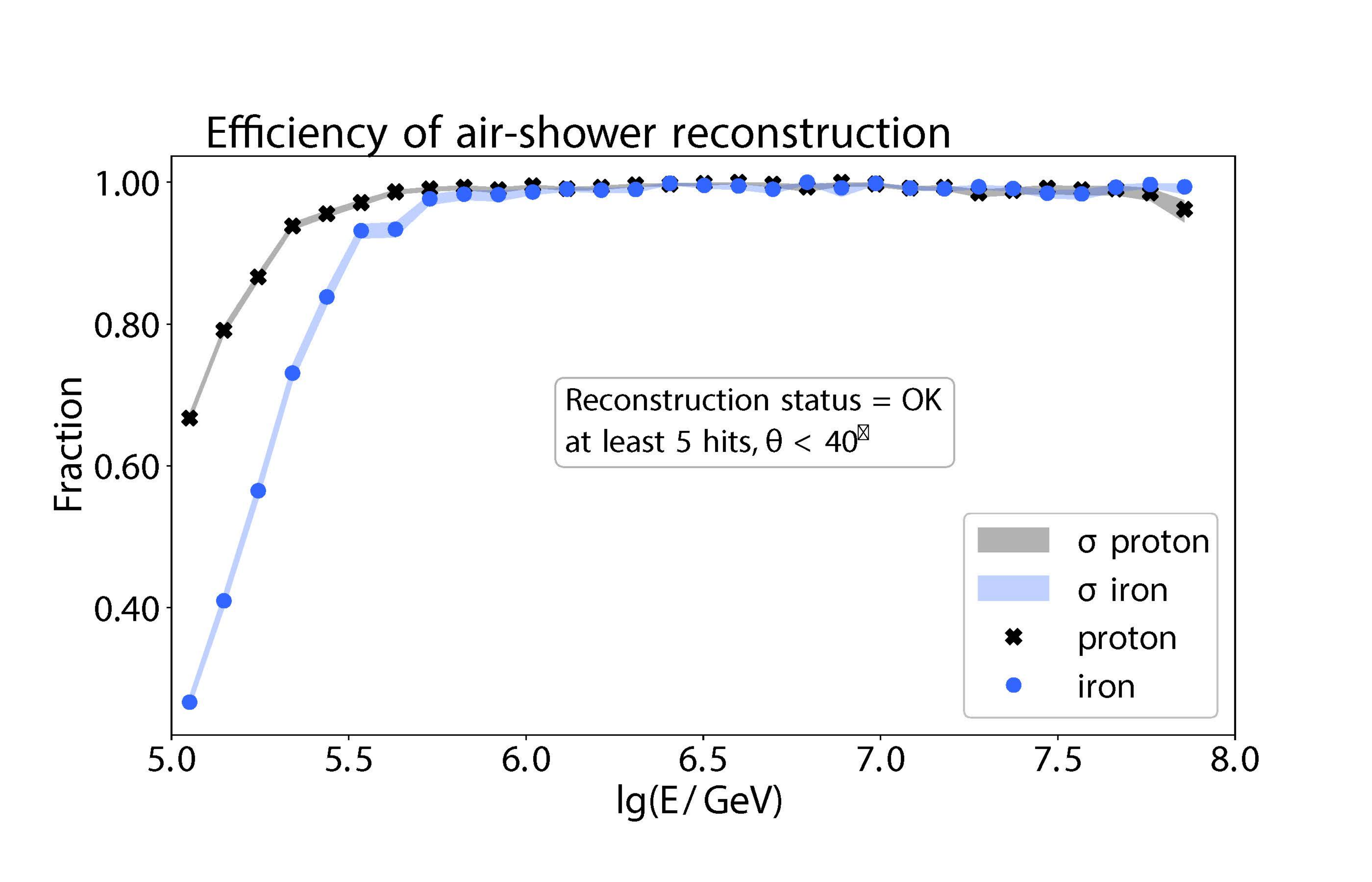}
\caption{The efficiency of reconstructing air-showers for proton
and iron primaries. The proposed array can reconstruct air-showers
with full efficiency even below 1 PeV. Still high efficiency for lower
energies will improve the veto analysis for in-ice measurements.}
\label{fig-8}       
\end{figure}

The reconstruction is performed iteratively in three steps. 
The distribution of arrival delays with respect to the
plane-front is described by the sum of a parabolic and a Gaussian function per single event. 
Within the reconstruction the negative log-likelihood is minimized, taking the Poisson distribution for
non-zero signals and its error for silent stations, and Gaussian distribution for time delays.
The overall reconstruction shown in Figure~\ref{fig-8} was performed using the modified Linsley function
with parametrisation on one of the slope parameters~\cite{agnieszkaecrs}. It shows full efficiency for air-showers
even below 1 PeV, which can significantly improve the detailed studies of the transition region.
The average accuracy of reconstruction above 1 PeV is smaller than 1 degree in direction  and
around 12 m in core position. 
In summary, with such an array the detection and reconstruction of showers are possible with an efficiency larger 
than 50\% for energies above $10^{14.2}\,$eV. Thus, the proposed array will fulfill all science
requirements.
\begin{figure}[h]
\centering
\includegraphics[width=0.9\columnwidth,clip]{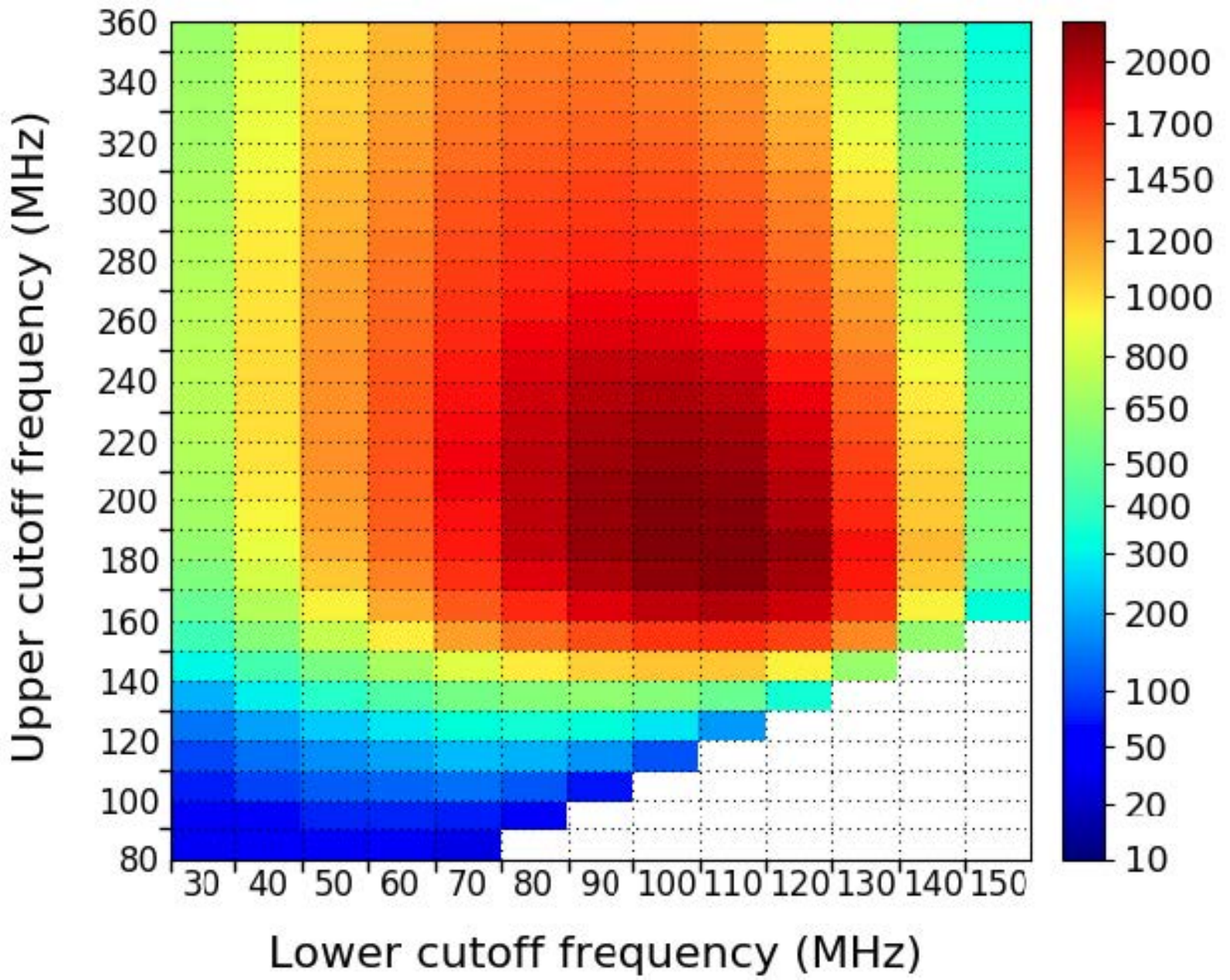}
\caption{Signal-to-noise behaviour at various frequency bands, for one typical shower induced by
a 10 PeV gamma-ray primary with zenith angle of 61 degree. It represents the SNR including the realistic noise situation at the South Pole plus a thermal noise of 
$40\,$K (SKALA antenna).}
\label{fig-9}       
\end{figure}

\section{Simulations of a radio antenna array}

The technique of air-shower detection via its radio emission has been 
successfully used in many experiments for the study of cosmic rays~\cite{frank-review}. 
However, an efficient detection was only possible for energies well above 
$10^{16.5}\,$eV. To use the technique efficiently at the South Pole to reach the science goals of the 
enhancement we focused on the possibility to lower the threshold by at least an order of magnitude~\cite{aswathiejpc,aswathiarena}. 
This is of particular importance to detect PeV gamma-rays 
from the Galactic Center visible all the time at 61 degree zenith angle.

\begin{figure}[h!]
\centering
\includegraphics[width=0.95\columnwidth,clip]{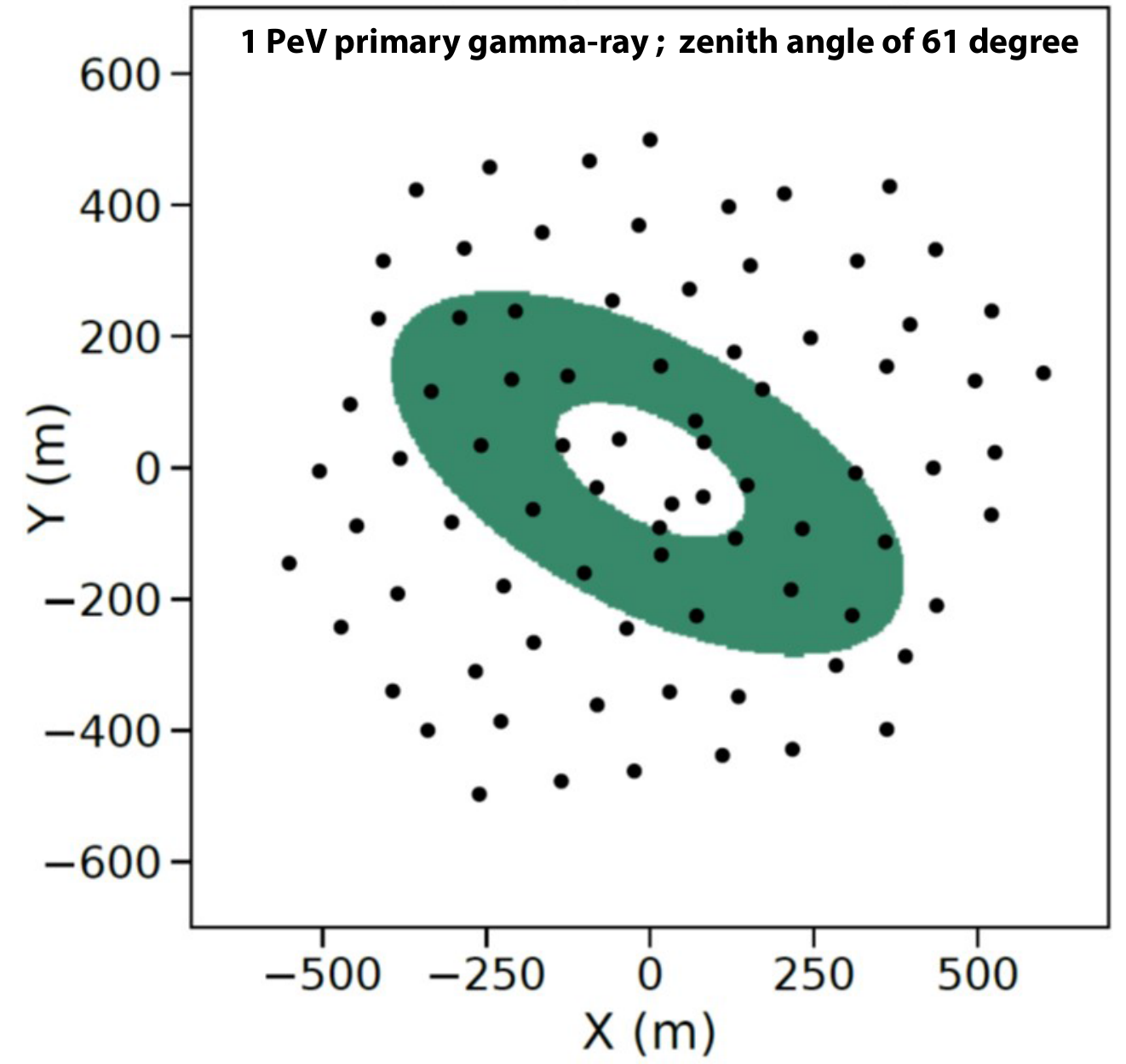}
\caption{Footprint of a 1 PeV gamma-ray shower at 100-190 MHz. 
The black dots represent the 81 antenna positions. Green represents an amplitude which is 
easily detectable by the SKALA antenna. Only the Cherenkov ring is
visible at this energy. It is sufficient to have three antennas within
the Cherenkov ring, in order to detect and reconstruct the shower.}
\label{fig-10}       
\end{figure}

One of the major challenges for the detection of air-showers with PeV energies with the radio detection
technique is the relatively weak radio signal in comparison to the noise, at these energies. This is
especially the case for the typical band of 30-80 MHz. It was shown that a higher level of signal-to-noise 
ratio is achieved by shifting the operational frequency band to 100-190 MHz (Figure~\ref{fig-9}). 
Figure~\ref{fig-10} shows the radio footprint of a gamma-ray shower with energy 1 PeV
that falls on an antenna array with one antenna placed at each IceTop station. 
The simulations include realistic noise at the South Pole plus a thermal noise contribution from the antenna as well as full shower and radio emission simulation by CoREAS~\cite{coreas}. 
An example for an antenna system with the required low level of thermal noise 
is the SKALA with a thermal noise level of 40 K. The use of such an antenna will lower the threshold down to around $1\,$PeV~\cite{phd-ab}. \\
The IceTop surface array together with the
future enhancement of IceTop with scintillators will not be able to reconstruct 
PeV air-showers from the direction of the Galactic Center, due to their low 
number of particles at the observation level. 
However, radio signals from these showers will survive and can thus be 
used for their observation. 
The studies have shown that the technique of radio detection of air showers, 
which has so far been used for the detection of cosmic rays, 
can also be used to observe PeV gamma-rays approaching the Earth from the Galactic 
Center.
This can be achieved by deploying an antenna array operating at the optimal frequency band 
of 100-190 MHz at the South Pole, thereby enabling the detection of air showers at much 
lower energies than that of the showers detected with the radio technique so far.
Further physics potential of such a radio array is described in~\cite{ARENA2018}.

\section{Summary}

An upgrade of the present IceCube surface array (IceTop) with scintillation detectors and possibly radio antennas is foreseen to be deployed in coming years. 
The work and construction of prototypes is in full swing and is starting to show very promising performance in the laboratory as well as at the South Pole. 
The full hybrid surface array of ice-Cherenkov detectors, scintillators and radio antennas 
will allow us to calibrate the impact of snow accumulation on the reconstruction of cosmic-ray showers detected by IceTop as well as improve the veto capabilities of the surface array. In addition, simulation studies predict a rich physics program of this upgrade such as an enhanced accuracy to mass composition of cosmic rays, search for PeV photons from the Galactic Center, or more thorough tests of the hadronic interaction models. \\ 

%
%
%


\end{document}